\documentstyle[12pt]{article}
\newcommand{\be}{\begin{eqnarray}}
\newcommand{\ee}{\end{eqnarray}}

\def\lsim{\mathrel{\rlap{\lower4pt\hbox{\hskip1pt$\sim$}}
    \raise1pt\hbox{$<$}}}               
\def\gsim{\mathrel{\rlap{\lower4pt\hbox{\hskip1pt$\sim$}}
    \raise1pt\hbox{$>$}}}               

\input{epsfig.sty}

\begin{document}

\Huge{\noindent{Istituto\\Nazionale\\Fisica\\Nucleare}}

\vspace{-3.9cm}

\Large{\rightline{Sezione SANIT\`{A}}}
\normalsize{}
\rightline{Istituto Superiore di Sanit\`{a}}
\rightline{Viale Regina Elena 299}
\rightline{I-00161 Roma, Italy}

\vspace{0.65cm}

\rightline{INFN-ISS 97/2}
\rightline{April 1997}

\vspace{1.5cm}

\begin{center}

\LARGE{Rare decays $B \to (K, K^*) (\ell^+ \ell^-, \nu \bar{\nu})$\\ in
the quark model}\footnote{{\bf To appear in Physics Letters B.}}\\

\vspace{1cm}

\large{D. Melikhov$^{(*)}$, N. Nikitin$^{(*)}$ and S. Simula$^{(**)}$}\\

\vspace{0.25cm}

\normalsize{$^{(*)}$Nuclear Physics Institute, Moscow State University\\
Moscow, 119899, Russia\\ $^{(**)}$Istituto Nazionale di Fisica Nucleare,
Sezione Sanit\`{a}\\ Viale Regina Elena 299, I-00161 Roma, Italy}

\end{center}

\vspace{0.5cm}

\begin{abstract}

\noindent Long-distance effects in exclusive rare semileptonic transitions
$B \to (K, K^*)$ are analysed within a relativistic quark model. The
meson transition form factors, describing the meson amplitudes of the
effective weak Hamiltonian, are calculated within the dispersion
formulation of the quark model as relativistic double spectral
representations through the wave functions of the initial and final
mesons. The dilepton spectra and lepton asymmetries are considered within
the framework provided by the Standard Model. It is found that, while the
non-resonant decay rates are very sensitive to quark model parameters,
the model dependence of the predicted dilepton forward-backward and
lepton polarization asymmetries is remarkably small, providing only an
overall $\sim 10 \%$ uncertainty. 

\end{abstract}

\vspace{0.5cm}


\vspace{0.25cm}


\newpage

\pagestyle{plain}

\indent The investigation of rare semileptonic decays of the $B$ meson
induced by the flavour-changing neutral current transitions $b \to (s, d)$
provides an important test of the Standard Model ($SM$) and its possible
extensions. Rare decays are forbidden at tree level and occur at the
lowest order only through one-loop diagrams. This fact opens the
possibility to probe at comparatively low energies the structure of the
theory at large mass scales, thanks to the contributions of virtual
particles in the loops. However, in order to reliably separate
(perturbative) short-distance effects, the (non-perturbative)
long-distance contributions, entering the amplitudes of exclusive rare
decays, should be known with enough accuracy. The theoretical
investigation of these contributions encounters the problem of describing
the hadron structure, which yields the main uncertainty in the predictions
of exclusive rare decays. In this letter exclusive rare semileptonic
transitions $B \to (K, K^*)$ are analysed within the framework of the $SM$
and adopting a relativistic constituent quark ($CQ$) model. It is shown
that the differential decay rates are very sensitive to quark model
parameters, whereas the model dependence of the predicted dilepton
forward-backward and lepton polarization asymmetries is quite small,
providing only an overall $\sim 10 \%$ uncertainty. 

\indent {\bf 1. Operator basis.} The effective weak Hamiltonian, which
describes the $b \to s \ell^+ \ell^-$ transition, has the following form
\cite{gws}
 \be
    \label{heff}
    {\cal H}_{eff} = - \frac{4G_F}{\sqrt{2}} V_{tb} V_{ts}^* ~ \sum_i
     C_i(\mu) ~ O_i(\mu)
 \ee
where $G_F$ is the universal Fermi constant, the quantities $C_i(\mu)$ are
the Wilson coefficients, obtained after integrating out the heavy
particles, and the $O_i$'s are the basis operators; within the $SM$, the
operators providing the main contribution to rare decays are
\cite{burasmuenz,ali}
 \be
    \label{basis}
     O_1 & = & \frac14 \left( \bar{s}_\alpha \gamma^\mu (1 - \gamma_5) 
     b_\alpha \right) \left( \bar{c}_\beta \gamma_\mu (1 - \gamma_5)
     c_\beta \right), \nonumber \\
     O_2 & = & \frac14 \left( \bar{s}_\alpha \gamma^\mu (1 - \gamma_5) 
     b_\beta \right) \left( \bar{c}_\beta \gamma_\mu (1-\gamma_5) c_\alpha
     \right), \nonumber \\
     O_7 & = & \frac{e}{32 \pi^2} \bar{s}_\alpha \sigma_{\mu \nu}
     [m_b (1 + \gamma_5) + m_s (1 - \gamma_5)] b_\alpha ~ F^{\mu \nu},
     \nonumber \\ 
     O_9 & = & \frac{e^2}{32 \pi^2} (\bar{s}_\alpha \gamma^\mu (1 -
     \gamma_5) b_\alpha) ~ (\bar{\ell} \gamma_\mu \ell), \nonumber \\
     O_{10} & = & \frac{e^2}{32\pi^2} (\bar{s}_\alpha \gamma^\mu (1 -
     \gamma_5) b_\alpha) ~ (\bar{\ell} \gamma_\mu \gamma_5 \ell),
  \ee

\noindent In Eq. (\ref{heff}) the renormalization scale $\mu$ is usually
chosen to be $\mu \simeq m_b$ in order to avoid large logarithms in the
matrix elements of the operators $O_i$. The Wilson coefficients $C_i$
reflect the specific features of the theory at large mass scales; they
are calculated at the scale $\mu \simeq M_W$ and then evolved down to
$\mu = m_b$ by the renormalization group equations. The analytic
expressions for $C_i(\mu)$ in the $SM$ can be found, e.g., in
\cite{burasmuenz}. In what follows, the values of the Wilson coefficients
at the scale $\mu = m_b = 5 ~ GeV$ are \cite{ali}: $C_1(m_b) = -0.235$,
$C_2(m_b) = 1.1$, $C_7(m_b) = -0.333$, $C_9(m_b) = 4.09$ and $C_{10}(m_b)
= -4.32$.

\indent The four-quark operators $O_1$ and $O_2$ generate both short- and
long-distance contributions to the effective weak Hamiltonian
(\ref{heff}). Both contributions can be taken into account by replacing
$C_9(m_b)$ with an effective coefficient $C_9^{eff}(m_b, q^2)$ given by
\cite{ali}
 \be
    \label{c9eff}
    C_9^{eff}(m_b, q^2) & = & C_9(m_b) + \left[3 C_1(m_b) + C_2(m_b)
    \right] \cdot \nonumber \\
    & & \left[h({m_c \over m_b}, {q^2 \over m_b^2}) +
    \frac{3}{\alpha_{em}^2} \kappa \sum_{V_i = J/\psi, \psi', ...}
    \frac{\pi \Gamma(V_i \to \ell \ell) M_{V_i}}{M_{V_i}^2 - q^2 - i
    M_{V_i} \Gamma_{V_i}} \right]
 \ee
where $q^2$ is the invariant mass squared of the lepton pair. The
short-distance contributions are contained in the function $h(m_c/m_b,
q^2/m_b^2)$, which describes the one-loop matrix element of the four-quark
operators $O_1$ and $O_2$ (see, e.g., \cite{burasmuenz} for its explicit
expression). The long-distance contribution, related to the formation of
intermediate $c \bar{c}$ bound states, is usually estimated by combining
the factorization hypothesis and the Vector Meson Dominance ($VMD$)
assumption \cite{ali,rescontr}; phenomenological analyses \cite{rescontr}
suggest that in order to reproduce correctly the branching ratio
$\mbox{BR}(B \to J / \psi X \to \ell^+ \ell^- X) =$ $\mbox{BR}(B \to J /
\psi X) ~ \cdot$ $\mbox{BR}(J / \psi \to \ell^+ \ell^-)$ the fudge factor
$\kappa$, which appears in Eq. (\ref{c9eff}) to correct
phenomenologically for inadequacies of the factorization + $VMD$
framework, should satisfy the approximate relation: $\kappa ~ \left[ 3
C_1(m_b) + C_2(m_b) \right] \approx 1$. To sum up, the effective weak
Hamiltonian has the following structure (cf. \cite{burasmuenz,ali,nunu})
 \be
    {\cal{H}}_{eff}(b \to s \ell^+ \ell^-) & = & {\frac{G_F}{\sqrt2}}
    {\frac{\alpha_{em}}{2\pi}} V_{ts}^* V_{tb} \left[ -2i 
    {\frac{m_b}{q^2}} C_7(m_b) (\bar{s} \sigma_{\mu \nu} q^{\nu} (1 +
    \gamma_5) b) (\bar{\ell} \gamma^{\mu} \ell) + \right. \nonumber \\
    & & \left. C_9^{eff}(m_b) (\bar{s} \gamma_{\mu} (1 - \gamma_5) b)
    (\bar{\ell} \gamma^{\mu} \ell) + C_{10}(m_b) (\bar{s} \gamma_{\mu}
    (1 - \gamma_5) b) (\bar{\ell} \gamma^{\mu} \gamma_5 \ell) \right],
    \nonumber \\
    \label{Heffrare}
    {\cal{H}}_{eff}(b \to s \nu \bar{\nu}) & = & \frac{G_F}{\sqrt{2}}
    \frac{\alpha_{em}}{2\pi \sin^2{\theta_W}} V_{tb} V^{*}_{ts} X(x_t) ~
    (\bar{s} \gamma_{\mu} (1 - \gamma_5) b) ~ (\bar{\nu} \gamma^{\mu} (1
    - \gamma_5) \nu)
 \ee
where $x_t = (m_t/M_W)^2$ and $X(x_t)$ is given in \cite{nunu}. At $m_t =
176 ~ GeV$ one has $X(x_t) = 2.02$. 

\indent {\bf 2. Meson form factors.} The long-distance contribution to $B
\to (K, K^*)$ decays is contained in the meson matrix elements of the
bilinear quark currents appearing in ${\cal{H}}_{eff}$, i.e. in the
relativistic invariant transition form factors of the vector,
axial-vector and tensor currents\footnote{In rare semileptonic decays
there is another long-distance effect, known as the weak annihilation,
which is caused by the Cabibbo-suppressed part of the four-fermion
operators not included in the operator basis (\ref{heff}). However, the
impact of this process in $B \to (K, K^*)$ transitions is negligible
\cite{ali}.}. The amplitudes of meson decays are induced by the quark
transition $b \to s$ through the vector $V_{\mu} = \bar{s} \gamma_{\mu}
b$, axial-vector  $A_{\mu} = \bar{s} \gamma_{\mu} \gamma^5 b$ and tensor
$T_{\mu\nu} = \bar{s} \sigma_{\mu\nu} b$ currents, with the following
covariant structure \cite{iw}
 \be
    \label{amplitudes}
    <P(M_2, p_2)| V_\mu(0) |P(M_1, p_1)> & = & f_+(q^2) ~ P_{\mu} +
    f_-(q^2) ~ q_{\mu}, \nonumber \\
    <V(M_2, p_2, e)| V_\mu(0) |P(M_1, p_1)> & = & 2g(q^2) ~
    \epsilon_{\mu\nu\alpha\beta} ~ e^{*\nu} ~ p_1^{\alpha} ~
    p_2^{\beta}, \nonumber \\
    <V(M_2, p_2, e)| A_\mu(0) |P(M_1, p_1)> & = & i e^{*\alpha} ~ [f(q^2)
    ~ g_{\mu\alpha} + a_+ (q^2) ~ p_{1\alpha} ~ P_{\mu} + a_-(q^2) ~
    p_{1\alpha} ~ q_{\mu}], \nonumber \\
    <P(M_2,p_2)| T_{\mu\nu}(0) |P(M_1, p_1)> & = & -2i ~ s(q^2) ~
    (p_{1\mu} p_{2\nu} - p_{1\nu} p_{2\mu}), \nonumber \\
    <V(M_2, p_2, e)| T_{\mu\nu}(0) |P(M_1, p_1)> & = & ie^{*\alpha} ~
    [g_+(q^2) ~ \epsilon_{\mu\nu\alpha\beta} ~ P^{\beta} + g_-(q^2) ~
    \epsilon_{\mu\nu\alpha\beta} ~ q^{\beta} + \nonumber \\
    & & h(q^2) ~ p_{1\alpha} ~ \epsilon_{\mu\nu\beta\gamma} ~ p_1^{\beta}
    ~ p_2^{\gamma}],
    \ee
where $q = p_1 - p_2$ and $P = p_1 + p_2$. 

\indent The relativistic invariant form factors, appearing in Eq.
(\ref{amplitudes}), contain information on the non-perturbative aspects of
the decay processes, so that they should be calculated within a
non-perturbative approach for any particular initial and final mesons. To
this end various theoretical methods have been adopted, like: the
light-cone quark model ($LCQM$) \cite{jauswyler}, the constituent quark
picture \cite{stech}, the heavy-quark symmetry ($HQS$) relations
\cite{burdman}, the three-point sum rules ($3pSR$) \cite{colangelo} and
the light-cone sum rules ($LCSR$) \cite{aliev}. The main outcome of
existing analyses is the remarkable model-dependence of the predicted
form factors, illustrated in Table 1 in terms of another set of
frequently used form factors, namely: $F_1(q^2) = f_+(q^2)$, $F_0(q^2) =
f_+(q^2) + q^2 f_-(q^2) / (Pq)$, $F_T(q^2) = -(M_1 + M_2) s(q^2)$,
$V(q^2) = (M_1 + M_2) g(q^2)$, $A_1(q^2) = f(q^2) / (M_1 + M_2)$,
$A_2(q^2) = -(M_1 + M_2) a_+(q^2)$, $A_0(q^2) = [q^2 a_-(q^2) + f(q^2) +
(Pq) a_+(q^2)] / 2M_2$, $T_1(q^2) = -g_+(q^2)$, $T_2(q^2) = -g_+(q^2) -
q^2 g_-(q^2) / (Pq)$ and $T_3(q^2) = (M_1 + M_2)^2 [g_-(q^2) / (Pq) -
h(q^2) / 2]$. This fact causes quite uncertain predictions for branching
ratios and, in particular, for dilepton spectra and asymmetries within
the $SM$ \cite{ali,jauswyler,burdman,colangelo,aliev,gengkao}. It is
clear that such an uncertainty may become an obstacle for extracting
information on the Wilson coefficients (particularly, their signs) and
for analysing possible deviations from $SM$ predictions, like those
expected in $SUSY$ models \cite{ali,giw}. 

\indent We have investigated the relevant meson form factors within a
relativistic $CQ$ model adopting a dispersion formulation, which has
proved to be successful in describing semileptonic decays of heavy mesons
\cite{m}. The dispersion formulation of the quark model has several
contact points with the $LCQM$ (see, e.g., \cite{jaus}). However, the
$LCQM$ has the problem of a direct application at time-like values of
$q^2$ because of the contribution arising from the so-called non-partonic
diagram, which cannot be killed at $q^2 > 0$ by an appropriate choice of
the reference frame. The dispersion formulation overcomes this difficulty.
Indeed, the $LCQM$ form factors at $q^2 < 0$ are re-written as double
spectral representations in the invariant masses of the initial and final
$q \bar{q}$ pairs, and, then, an analytical continuation is performed to
reach the time-like region $q^2 > 0$.

\indent Let us consider the transition from the initial meson $q(m_2)
\bar{q}(m_3)$ with mass $M_1$ to the final meson $q(m_1) \bar q(m_3)$
with mass $M_2$, induced by the quark transition $m_2 \to m_1$ through
the current $\bar{q}(m_1) J_{\mu (\nu)} q(m_2)$. For the transition $B_u
\to (K, K^*)$ one has $m_2 = m_b$, $m_1 = m_s$ and $m_3 = m_u$. The
$CQ$ structure of the initial and final mesons is described by the
vertices $\Gamma_1$ and $\Gamma_2$, respectively. The initial $B$-meson
vertex has the spinorial structure $\Gamma_1 = i\gamma_5 ~ G_1 /
\sqrt{N_c}$, where $N_c$ is the number of colours; the final meson vertex
has the structure $\Gamma_2 = i\gamma_5 ~ G_2 / \sqrt{N_c}$ for a
pseudoscalar state and $\Gamma_{2\mu} = [A \gamma_\mu + B(k_1 - k_3)_\mu]
~ G_2 / \sqrt{N_c}$, with $A = -1$ and $B = 1 / (\sqrt{s_2} + m_1 + m_3)$
for an $S$-wave vector meson. At $q^2 < 0$ the form factors are given by
the following spectral representation \cite{m}
 \be
    \label{ff}
    f_i(q^2) = \int\limits^\infty_{(m_1 + m_3)^2} \frac{ds_2 ~
    G_2(s_2)}{\pi(s_2 - M_2^2)} \int\limits^{s_1^+(s_2, q^2)}_{s_1^-(s_2,
    q^2)} \frac{ds_1 ~ G_1(s_1)}{\pi(s_1 - M_1^2)} \frac{\tilde{f}_i(s_1,
    s_2, q^2)}{16 \lambda^{1/2}(s_1, s_2, q^2)}
 \ee
where $s_1^\pm(s_2, q^2) \equiv$ $[s_2 (m_1^2 + m_2^2 - q^2) + q^2 (m_1^2
+ m_3^2) - (m_1^2 - m_2^2) (m_1^2 - m_3^2) ~ \pm$ $\lambda^{1/2}(s_2,
m_3^2, m_1^2) $ $\lambda^{1/2}(q^2, m_1^2, m_2^2)] / 2m_1^2$ and
$\lambda(s_1, s_2, s_3) \equiv$ $(s_1 + s_2 - s_3)^2 - 4 s_1 s_2$ is the
triangle function. Equation (\ref{ff}) corresponds only to the
contribution of the two-particle singularities in the Feynman graphs. For
pseudoscalar and vector mesons with mass $M$, built up of $CQ$'s with
masses $m_q$ and $m_{\bar{q}}$, the function $G(s)$ can be written as 
 \be
    \label{vertex}
    G(s) = \frac{\pi}{\sqrt{2}} \frac{\sqrt{s^2 - (m_q^2 -
    m_{\bar{q}}^2)^2}} {\sqrt{s - (m_q - m_{\bar{q}})^2}} \frac{s -
    M^2}{s^{3/4}} w(k^2)
 \ee
where $k = \lambda^{1/2}(s, m_q^2, m_{\bar{q}}^2) / 2 \sqrt{s}$ and
$w(k^2)$ is the ground-state $S$-wave radial wave function, normalized as
$\int_0^{\infty} dk k^2 |w(k^2)|^2 = 1$. The double spectral densities
$\tilde{f}_i(s_1, s_2, q^2)$, appearing in Eq. (\ref{ff}), are explicitly
given by
 \be
    \label{ffs}
    \tilde{f}_+ + \tilde{f}_- & = & 4 [m_1 m_2 \alpha_1 - m_2 m_3 \alpha_1
    + m_1 m_3 (1 - \alpha_1) - m_3^2 (1 - \alpha_1) + \alpha_2 s_2],
    \nonumber \\
    \tilde{f}_+ - \tilde{f}_- & = & 4 [m_1 m_2 \alpha_2 - m_1 m_3
    \alpha_2 + m_2 m_3 (1 - \alpha_2) - m_3^2 (1 - \alpha_2) + \alpha_1
    s_1], \nonumber \\
    \tilde{g} & = & -2A ~ [m_1\alpha_2 + m_2 \alpha_1 + m_3 (1 - \alpha_1
    - \alpha_2)]- 4 B \beta, \nonumber \\  
    \tilde{a}_+ + \tilde{a}_- & = & -4A [2m_2 \alpha_{11} + 2m_3 (\alpha_1
    - \alpha_{11})] + 4B [C_1 \alpha_1 + C_3 \alpha_{11}], \nonumber \\ 
    \tilde{a}_+ - \tilde{a}_- & = & -4A [-m_1 \alpha_{2} - m_2 (\alpha_{1}
    - 2\alpha_{12}) - m_3 (1 - \alpha_1 - \alpha_2 + 2\alpha_{12})] + 4B
    [C_2 \alpha_1 + C_3 \alpha_{12}], \nonumber \\ 
    \tilde{f} & = & \tilde{f}_D + (M_1^2 - s_1 + M_2^2 - s_2) \tilde{g},
    \nonumber \\
    \tilde{s} & = & 2 [m_1 \alpha_2 + m_2 \alpha_1 + m_3 (1 - \alpha_1 -
    \alpha_2)], \nonumber \\
    \tilde{g}_+ + \tilde{g}_- & = & 4A [m_3 (m_1 - m_3) + \alpha_1 (m_1
    - m_3) (m_2 - m_3) + \alpha_2 s_2 + 2\beta] + 8B (m_1 + m_3) \beta,
    \nonumber \\
    \tilde{g}_+ - \tilde{g}_- & = & 4A [m_3 (m_2 - m_3) + \alpha_2 (m_1 -
    m_3) (m_2 - m_3) + \alpha_1 s_1] + 8B (m_2 - m_3) \beta, \nonumber \\ 
    \tilde{h} & = & -8A \alpha_{12} - 8B [-m_3 \alpha_1 + (m_3 - m_2)
    \alpha_{11} + (m_3 + m_1) \alpha_{12}],
 \ee
where $\tilde{f}_D = -4A [m_1 m_2 m_3 + m_2 (s_2 - m_1^2 - m_3^2) / 2 +
m_1 (s_1 - m_2^2 - m_3^2) / 2 -$ $m_3 (s_3 - m_1^2 - m_2^2) / 2 +
2\beta (m_2 - m_3)] + 4B C_3 \beta$, $\alpha_1 = [(s_1 + s_2 - s_3) (s_2 -
m_1^2 + m_3^2) - 2s_2 (s_1 - m_2^2 + m_3^2)] / \lambda(s_1, s_2, s_3)$,
$\alpha_2 = [(s_1 + s_2 - s_3) (s_1 - m_2^2 + m_3^2) - 2s_1 (s_2 - m_1^2
+ m_3^2)] / \lambda(s_1, s_2, s_3)$, $\beta = [2m_3^2 - \alpha_1 (s_1 -
m_2^2 + m_3^2) - \alpha_2 (s_2 - m_1^2 + m_3^2)] / 4$, $\alpha_{11} = 
\alpha_1^2 + 4\beta(s_2) / \lambda(s_1, s_2, s_3)$, $\alpha_{12} =
\alpha_1 \alpha_2 - 2 \beta(s_1 + s_2 - s_3) / \lambda(s_1, s_2, s_3)$,
$C_1 = s_2 - (m_1 + m_3)^2$, $C_2 = s_1 - (m_2 - m_3)^2$ and $C_3 = s_3 -
(m_1 + m_2)^2 - C_1 - C_2$.

\indent At $q^2 < 0$ the representation (\ref{ff}) with the spectral
densities (\ref{ffs}) for all the form factors but $f(q^2)$ coincide with
the corresponding $LCQM$ expressions (see, e.g., \cite{jaus}). This is
due to the fact that within the dispersion approach all the form factors
but $f(q^2)$ are given by a double dispersion representations without
subtractions, whereas in order to construct the form factor $f(q^2)$ from
its double spectral density a subtraction procedure is necessary. We fix
this procedure by requiring that, in case of meson transitions induced by
a heavy-to-heavy quark transition, all the form factors, including
$f(q^2)$, satisfy the leading-order Isgur-Wise ($IW$) relations
\cite{iw} as well as the subleading $O(1 / m_Q)$ relations of the $1 /
m_Q$ expansion \cite{luke}, which are model-independent consequences of
$QCD$. This is fulfilled  by the form factors given by Eq. (\ref{ff})
with the spectral densities (\ref{ffs}), provided that the functions
$G_i(s_i)$ are localized near the $q \bar{q}$ threshold with a width of
the order $\Lambda_{QCD}$ \cite{mn}. Moreover, for meson decays induced
by a heavy-to-light quark transition the dispersion formulation satisfies
the leading-order relations between the form factors of the vector and
tensor currents given in \cite{iw}. 

\indent The analytical continuation to the time-like region $q^2 > 0$
generates two contributions: the first one is the normal contribution,
which is just the expression (\ref{ff}) taken at $q^2 > 0$, and the
second one is  an additional anomalous contribution, described explicitly
in \cite{m}. The normal contribution dominates the form factors at small
$q^2$ and vanishes when $q^2 = (m_2 - m_1)^2$, while the anomalous
contribution is negligible at small $q^2$ and steeply rises as $q^2 \to
(m_2 - m_1)^2$.

\indent {\bf 3. Results.} We ran calculations adopting two different
$CQ$ models, which will be referred to as $Set ~ 1$ and $Set ~ 2$. In the
former the simple Gaussian ans\"{a}tz of the $ISGW2$ model \cite{isgw2}
is used for $w(k^2)$ in Eq. (\ref{vertex}), whereas in the latter model
the radial function $w(k^2)$ is the variational solution of the effective
$q \bar{q}$ semi-relativistic Hamiltonian of Godfrey and Isgur ($GI$)
\cite{gi}. The values of the $CQ$ masses as well as the average of the
internal momentum squared, $\langle k^2 \rangle$, are reported in Table
2. The main difference between the radial functions $w(k^2)$ in the two
$CQ$ models relies in their behaviour at high values of the internal
momentum $k$ (see Table 2 and cf. \cite{sim95,sim96}): while the Gaussian
ans\"{a}tz yields a soft wave function, which takes into account mainly
the effects of the confinement size, the $GI$ wave functions exhibit
high-momentum components generated by the effective one-gluon-exchange of
the $GI$ potential. The impact of these components on the predicted
universal $IW$ function and on the form factor $f_+(q^2)$ for
heavy-to-light transitions has been analyzed in \cite{sim96} and found
not negligible.

\indent The results of our calculations of the form factors have been
fitted in terms of a simple $q^2$ behaviour of the form $f_i(q^2) =
f_i(0) / [1 - \sigma_1 q^2 + \sigma_2 q^4]$; the resulting values of the
parameters $f_i(0)$, $\sigma_1$ and $\sigma_2$ are reported in Table 3. It
can clearly be seen that the form factors are very sensitive to the
choice of the quark model. In particular, the form factors obtained with
the $GI$ wave function ($Set ~ 2$) are systematically larger than those
corresponding to the Gaussian-like ans\"{a}tz ($Set ~ 1$). This feature
is related both to the larger content of high-momentum components and to
the lower values of the $CQ$ masses characterizing the model of Ref.
\cite{gi} with respect to the $ISGW2$ model (see Table 2).

\indent The behaviour of the form factors of interest is known in case of
heavy parent and daughter quarks inducing the heavy-to-heavy meson
transition $M_1 \to M_2$. The leading-order $1/m_Q$ relations read as
\cite{iw}
 \be
    \label{hh}
    f_+(\omega) & = & -g_+(\omega) = \frac{M_1 + M_2}{2\sqrt{M_1M_2}} ~
    \xi(\omega), ~~~~~~~~
    f_-(\omega) = -g_-(\omega) = - \frac{M_1-M_2}{2\sqrt{M_1M_2}} ~
    \xi(\omega), \nonumber \\
    s(\omega) & = & g(\omega) = a_-(\omega) = - a_+(\omega) =
    \frac{1}{2\sqrt{M_1M_2}} ~ \xi(\omega), \nonumber \\
    f(\omega) & = & \sqrt{M_1M_2} ~ (1+\omega) ~ \xi(\omega),
     ~~~~~~~~~~~~~~ h(\omega) = \frac{1}{\sqrt{M_1M_2}} ~ O(1/m_Q),
 \ee
where $\omega = (M_1^2 + M_2^2 - q^2) / 2M_1M_2$ and $\xi(\omega)$ is the
$IW$ function. One should not expect these relations to work with high
accuracy for the $B \to (K, K^*)$ transitions, because the daughter
$s$-quark cannot be considered heavy enough. As a matter of fact, using
the $IW$ function calculated in \cite{sim96}, we have checked that large
violations of Eq. (\ref{hh}) occur, especially far from the zero-recoil
point. However, Eq. (\ref{hh}) implies some (approximate) $HQS$ relations
among the form factors, namely: $F_1(q^2) \simeq V(q^2) \simeq A_0(q^2)
\simeq A_2(q^2) \simeq - F_T(q^2) \simeq T_1(q^2) \simeq T_3(q^2)$ and
$F_0(q^2) \simeq A_1(q^2) \simeq T_2(q^2)$, which turn out to be
fulfilled within $\sim 20 \%$ for both $Set ~ 1$ and $Set ~ 2$ in the
whole kinematical range.

\indent The averages of our predictions at $q^2 = 0$ obtained within $Set
~ 1$ and $Set ~ 2$ is reported in Table 1. It can be seen that our form
factors are consistent with the $LCSR$ results. At the same time, a
striking disagreement with some of the $3pSR$ results has been found,
especially in case of the form factors $F_T$ and $T_3$. Indeed, the form
factor $T_3$ of Ref. \cite{colangelo} has an opposite sign and is large
in absolute value compared with $T_1$, as well as the form factor $|F_T|$
is small compared with $F_1$, at variance with the approximate $HQS$
relations. We want to mention that in Ref. \cite{colangelo} both $F_T$
and $T_3$ are claimed to be constructed in terms of some of the other form
factors applying equations of motion. However, it can be checked that the
expression given in \cite{colangelo} for $T_3$ does not satisfy the $HQS$
requirements in the heavy-quark limit.

\indent The differential decay rates and asymmetries have been calculated
using the expressions given in Refs. \cite{colangelo,gengkao,giw}. The
predictions for the dilepton distribution in $B \to (K, K^*) \mu^+ \mu^-$
decays are reported in Fig. 1, where the non-resonant contributions are
shown separately. The total decay rates turn out to be at least one
order of magnitude larger than the non-resonant decay rates. However, the
resonant contributions are strongly peaked in narrow regions around their
masses, so that outside these regions the resonance influence is almost
negligible. This fact allows to reliably separate the resonant
contributions from the non-resonant one, which contains the relevant
information on the Wilson coefficients. In Table 4 our predictions for
the non-resonant decay rates and branching ratios are listed. The
dependence on the chosen quark model is strong, yielding a large
uncertainty in the predictions; such a drawback may be limited by testing
the same $CQ$ model in several semileptonic decays, including in
particular the shape of differential decay rates and lepton spectra,
which are expected to be sensitive to the form of the $CQ$ model wave
functions. Finally, note that the transitions $B \to K^* \mu^+ \mu^-$ and
$B \to K^* e^+ e^-$ have different rates, because the amplitude $B \to
K^* \ell^+ \ell^-$ has a kinematical pole at $q^2 = 0$, which makes the
corresponding decay rate very sensitive to the lower boundary of the
phase space volume ($q^2 = 4m_{\ell}^2$), while the amplitude $B \to K
\ell^+ \ell^-$ is regular at $q^2 = 0$ and, therefore, insensitive to the
mass of the light lepton. 

\indent Our results for total decay rates and branching ratios are
summarized in Table 5 and compared with the predictions of other
approaches. It can be seen that our results are consistent with those of
Ref. \cite{ali}, which are based on the application of the $HQS$ relations
for the form factors, whereas they differ from the  predictions of Ref.
\cite{colangelo}. On one hand, the present level of model dependence does
not provide the opportunity to extract precise values of $V_{ts}$ and to
study possible effects beyond the $SM$. On the other hand, the situation
with lepton asymmetries looks much more optimistic. Our results for the
dilepton forward-backward and lepton polarization asymmetries are
presented in Figs. 2 and 3. The asymmetries calculated within different
quark models turn out to differ only by a very small amount ($\sim 10
\%$) in the whole kinematical range. This is due to the following facts:
i) the asymmetries depend on ratios among form factors; ii) our form
factors obey approximate $HQS$ relations, which means that their ratios
are only slightly model-dependent. We point out that our dilepton
asymmetry is approximately a factor of $2$ lower than the result of Ref.
\cite{colangelo}\footnote{We have taken the resonance phase in accordance
with Ref. \cite{rescontr}, whereas an opposite sign is used in
\cite{colangelo}.} and this disagreement can be traced back again to the
large difference in the form factor $T_3$ calculated in \cite{colangelo}
and in the present work (see Table 1). 

\indent In conclusion, we have analysed rare semileptonic transitions $B
\to (K, K^*)$ within a relativistic constituent quark model, formulated in
a dispersion form. The differential decay rates are found to be strongly
sensitive to the choice of the particular quark model, obtaining an
overall $\sim 50 \%$ uncertainty in the predictions. However, at the same
time, the asymmetries of lepton distributions result to be almost
insensitive to quark model parameters, so that they can be predicted
within the framework of the Standard Model with little uncertainty ($\sim
10 \%$). Our predictions for lepton asymmetries in rare semileptonic 
decays may provide a reliable starting point for investigating possible
violations of Standard Model predictions, like those expected in $SUSY$
models.

\vspace{0.5cm}

\noindent {\bf Acknowldgments.} We are grateful to B. Stech and K.A.
Ter-Martirosyan for useful discussions and to L. Smirnova for her
interest in this work, which was supported by the $RFBR$ under grant
96-02-18121a.

\newpage

{\footnotesize

\begin{table}[htb]

\noindent {\bf Table 1.} Transition form factors for the semileptonic 
decays $B \to (K, K^*)$ evaluated at $q^2 = 0$ within various theoretical 
methods. The last row represents the average of the results obtained in 
the present work adopting the two $CQ$ models, $Set ~ 1$ ($ISGW2$ wave
function \cite{isgw2}) and $Set ~ 2$ ($GI$ wave function \cite{gi}),
described in the text.

\begin{center}

\begin{tabular}{||l||l|l||l|l|l||l|l||} \hline
$Ref.$ & $F_1(0)$ & $F_T(0)$& $V(0)$ & $A_1(0)$ & $A_2(0)$ &$T_1(0)$
       & $T_3(0)$ \\ \hline
$LCQM ~ \cite{jauswyler}$ & 0.30 & $-0.30$ & 0.35 & 0.26 & 0.24 & 0.31
       & 0.32\\  
$3pSR ~ \cite{colangelo}$ & $0.25\pm0.03$ & $-0.14\pm0.03$
       & $0.47\pm0.03$    & $0.37\pm0.03$ & $ 0.40\pm0.03$
       & $0.38\pm0.06$    & $-1.96$\\   
$LCSR ~ \cite{aliev}$     & 0.29 & $-0.31$ & 0.45 & 0.36 & 0.40 &0.36
       &0.39\\ \hline  
$This ~ work$             & $0.43\pm0.07$ & $-0.40\pm0.06$
       & $0.40\pm0.10$    & $0.34\pm0.08$ & $ 0.27\pm0.05$
       & $0.37\pm0.09$    & $0.35\pm0.09$\\ \hline  
\end{tabular}

\end{center}

\end{table}

\vspace{2cm}

\begin{table}[htb]

\noindent {\bf Table 2.} Values of the $CQ$ masses and of the average 
internal momentum squared $\langle k^2 \rangle$, in $(GeV/c)^2$, for the 
two $CQ$ models adopted in this work.

\begin{center}

\begin{tabular}{||c||c|c|c|c||c|c|c||} \hline
$Ref.$ & $m_u$ & $m_s$ & $m_c$ & $m_b$ & $<k^2>_K$ & $<k^2>_{K^*}$
       & $<k^2>_{B_u}$\\ \hline
$Set ~ 1 ~ \cite{isgw2}$ & 0.33 & 0.55 & 1.82 & 5.2 & 0.29 & 0.16 & 0.28\\
       \hline 
$Set ~ 2 ~ \cite{sim95}$ & 0.22 & 0.42 & 1.65 & 5.0 & 0.77 & 0.34 & 0.60\\
       \hline
\end{tabular}

\end{center}

\end{table}

\vspace{2cm}

\begin{table}[htb]

\noindent {\bf Table 3.} Parameters of the fit $f_i(q^2) = f_i(0) / [1 -
\sigma_1 q^2 + \sigma_2 q^4]$ to the $B \to (K, K^*)$ transition form
factors calculated within the two $CQ$ models adopted.

\begin{center}

\begin{tabular}{||c||c|c|c||c|c|c|c||c|c|c||} \hline
$Decays$  & \multicolumn{3}{c|}{$B \to K$} & \multicolumn{7}{c|}{$B \to
          K^*$}\\ \hline
          & $f_{+}(0)$ & $f_{-}(0)$ & $s(0)$ & $g(0)$   & $f(0)$
          & $a_+(0)$   & $a_{-}(0)$ & $h(0)$ & $g_+(0)$ & $g_-(0)$\\  
$Ref.$    &$\sigma_1$&$\sigma_1$&$\sigma_1$&$\sigma_1$&$\sigma_1$
          &$\sigma_1$&$\sigma_1$&$\sigma_1$&$\sigma_1$&$\sigma_1$\\
          &$\sigma_2$&$\sigma_2$&$\sigma_2$&$\sigma_2$&$\sigma_2$
          &$\sigma_2$&$\sigma_2$&$\sigma_2$&$\sigma_2$&$\sigma_2$\\
          \hline
$Set ~ 1$ & 0.36   & $-$0.30 & 0.06   & 0.048  & 1.60   & $-$0.036 
          & 0.041  & 0.0037  & $-$0.28& 0.24\\
          & 0.048  & 0.050   & 0.049  & 0.057  & 0.0288 &  0.053   
          & 0.055  & 0.075   & 0.058  & 0.059\\
          & 0.00063& 0.00061 & 0.00064& 0.00085& 0.00028& 0.00082
          & 0.00088& 0.0016  & 0.0009 & 0.00096\\ \hline 
$Set ~ 2$ & 0.50   &$-$0.42  & 0.080  & 0.083  & 2.62   & $-$0.052
          & 0.067  & 0.0085  &$-$0.47 & 0.43\\
          & 0.035  & 0.036   & 0.035  & 0.042  & 0.011  &  0.030
          & 0.040  & 0.044   & 0.042  & 0.043\\
          & 0.00017& 0.00018 & 0.00017& 0.00036& 0.00015& $-$0.00025
          & 0.00016& 0.00001 & 0.00036& 0.00037\\ \hline
\end{tabular}

\end{center}

\end{table}

\newpage

\begin{table}[htb]

\noindent {\bf Table 4.} Non-resonant decay rates and branching ratios of 
the decays $B \to (K, K^*)$ $(\ell^+ \ell^-, \nu \bar{\nu})$. Decay rates 
are given in units $|V_{ts}^* V_{tb}|^2 \times 10^{8} ~ s^{-1}$, while
branching ratios are in units $|V_{ts}^* V_{tb}|^2 \times 10^{-4}$.

\begin{center}

\begin{tabular}{||c c||c|c|c|c|c||} \hline
$Ref.$    & & $B \to K \ell^+ \ell^-$ & $B \to K \sum \nu_i \bar{\nu}_i$
            & $B \to K^* e^+ e^-$     & $B \to K^* \mu^+ \mu^-$
            & $B \to K^* \sum \nu_i \bar{\nu}_i$\\ \hline
$Set ~ 1$ & $\Gamma$     & 2.0 & 25 &  5.2 &  4.1 &  51\\   
          & ${\cal{BR}}$ & 2.9 & 37 &  7.6 &  6.0 &  75\\ \hline
$Set ~ 2$ & $\Gamma$     & 3.3 & 39 & 11.5 &  8.5 & 108\\  
          & ${\cal{BR}}$ & 4.8 & 57 & 17.0 & 12.5 & 160\\ \hline
\end{tabular}

\end{center}

\end{table}

\vspace{2cm}

\begin{table}[htb]

\noindent {\bf Table 5.} Non-resonant decay rates and branching ratios of 
rare radiative and semileptonic decays of $B$-meson. Theoretical 
predictions are given in units $|V_{ts} / 0.033|^2$. The results of Ref.
\cite{colangelo} have been recalculated replacing the value $|V_{ts} /
0.04|^2$ with $|V_{ts} / 0.033|^2$. The uncertainties in $V_{ts}$ are not
included in the error bars.

\begin{center}

\begin{tabular}{||l||c|c|c||c||} \hline
Ref.                    & This work & \cite{ali} & \cite{colangelo}
                        & Exp. \\ \hline
Decay                   & $\Gamma ~ (s^{-1})$ & $\Gamma ~ (s^{-1})$
                        & $\Gamma ~ (s^{-1})$ & $\Gamma ~ (s^{-1})$\\
mode                    & ${\cal{BR}}$        & ${\cal{BR}}$
                        & ${\cal{BR}}$        & ${\cal{BR}}$\\
                        \hline \hline
$B \to K^* \gamma$      & $(2.9 \pm 1.3) \times 10^7$    & $--$ & $--$
                        & $--$\\
                        & $(3.9 \pm 1.7) \times 10^{-5}$
                        & $(4.9 \pm 2.0) \times 10^{-5}$ & $--$
                        & $(4.2 \pm 1.0) \times 10^{-5}$
                        \cite{cleo1} \\ \hline \hline
$B \to K \ell^+ \ell^-$ & $(2.9 \pm 0.7) \times 10^5$    & $--$ & $--$
                        & $--$\\
                        & $(4.2 \pm 0.9) \times 10^{-7}$
                        & $(4.0 \pm 1.5) \times 10^{-7}$
                        & $2 \times 10^{-7}$
                        & $< 0.9 \times 10^{-5}$ \cite{cleo2}\\ \hline
$B \to K \sum \nu_{i} \bar{\nu}_i$ & $(3.5 \pm 0.8) \times 10^6$
                                   & $--$& $--$ & $--$\\
                                   & $(5.2 \pm 1.1) \times 10^{-6}$     
                                   & $(3.2 \pm 1.6) \times 10^{-6}$
                                   & $(1.6 \pm 0.4) \times 10^{-6}$
                                   & $--$\\ \hline \hline
$B \to K^* e^+ e^-$                & $(9.2\pm3.5)\times 10^5$
                                   & $--$ & $--$ & $--$\\
                                   & $(1.4 \pm 0.5) \times 10^{-6}$     
                                   & $(2.3 \pm 0.9) \times 10^{-6}$
                                   & $0.7 \times 10^{-6}$
                                   & $ <1.6 \times 10^{-5}$ \cite{cleo2}
                                   \\ \hline
$B \to K^* \mu^+ \mu^-$            & $(6.9 \pm 2.4) \times 10^5$
                                   & $--$ & $--$ & $--$\\
                                   & $(1.0 \pm 0.4) \times 10^{-6}$     
                                   & $(1.5 \pm 0.6) \times 10^{-6}$
                                   & $0.7 \times 10^{-6}$
                                   & $< 2.5 \times 10^{-5}$ \cite{cdf}
                                   \\ \hline
$B \to K^* \sum \nu_i \bar{\nu}_i$ & $(8.7 \pm 3.1) \times 10^6$      
                                   & $--$ & $--$ & $--$ \\
                                   & $(1.3 \pm 0.5)  \times 10^{-5}$     
                                   & $(1.1 \pm 0.55) \times 10^{-5}$
                                   & $(3.5 \pm 0.5)  \times 10^{-6}$
                                   & $--$\\ \hline
\end{tabular}

\end{center}

\end{table}

\newpage

\begin{figure}[htb]

\epsfxsize=16.5cm \epsfig{file=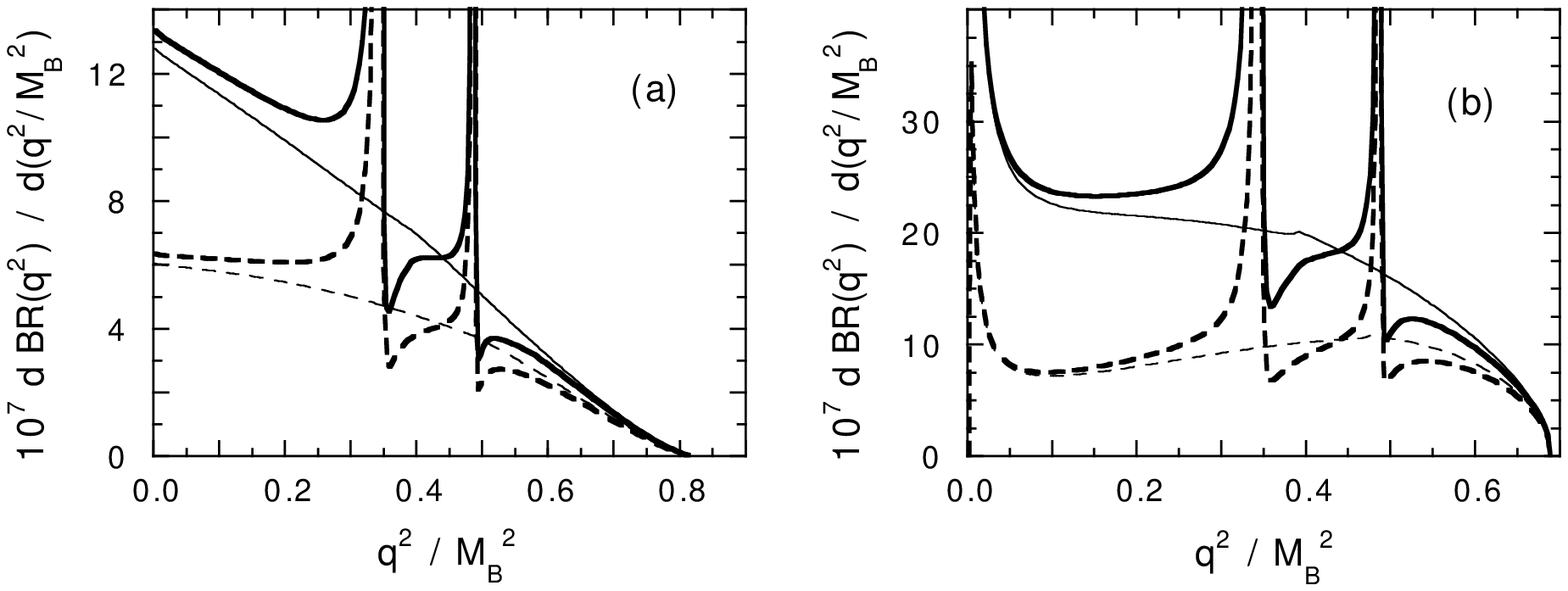}

\vspace{-17.5cm}

\noindent {\bf Figure 1.} Differential decay rate for the transitions $B 
\to K \ell^+ \ell^-$ (a) and $B \to K^* \mu^+ \mu^-$ (b) versus the 
invariant mass squared of the lepton pair, $q^2$, divided by the $B$-meson 
mass squared, $M_B^2$. Dashed and solid lines correspond to our calculations
obtained using the two $CQ$ models $Set ~ 1$ ($ISGW2$ wave function
\cite{isgw2}) and $Set ~ 2$ ($GI$ wave function \cite{gi}), respectively.
The thin lines represent the non-resonant contribution only, whereas the
thick lines are the sum of non-resonant and resonant contributions.

\end{figure}

\vspace{2cm}

\begin{figure}[htb]

\epsfxsize=16.5cm \epsfig{file=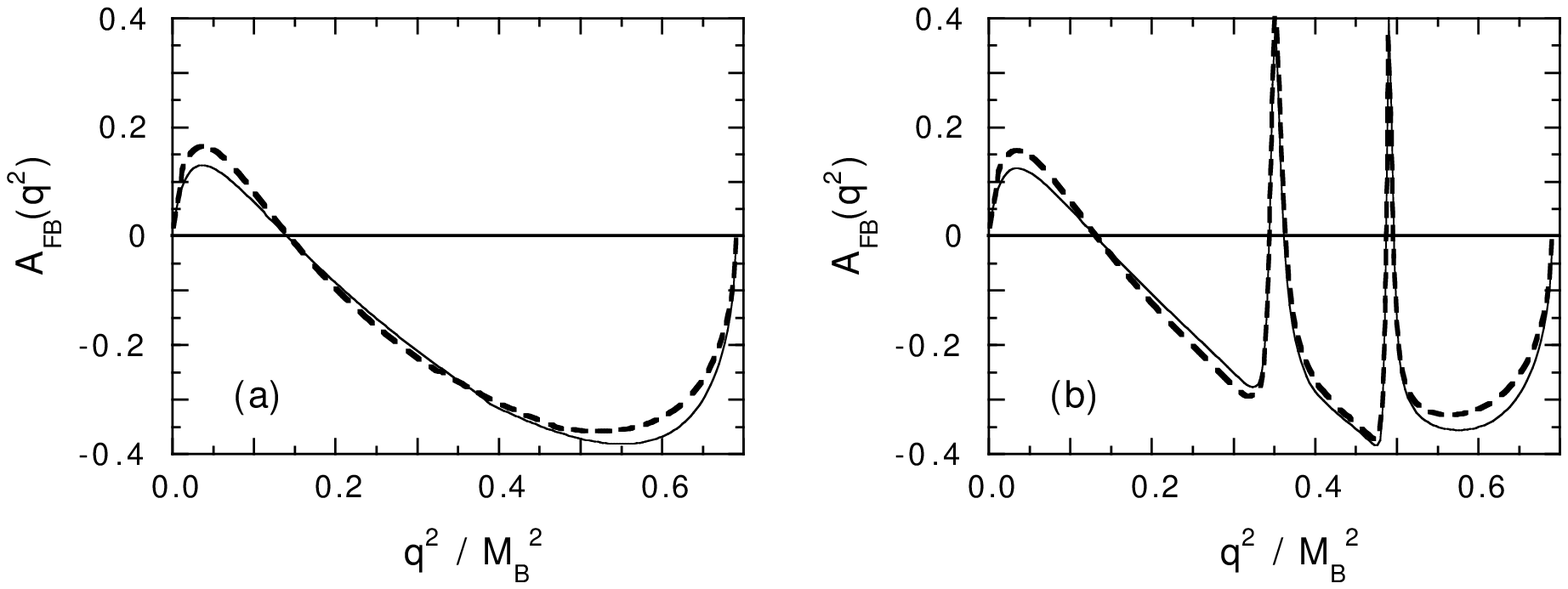}

\vspace{-17.5cm}

\noindent {\bf Figure 2.} The dilepton forward-backward asymmetry ($A_{FB}$)
for the decay $B \to K^* \mu^+ \mu^-$ versus $q^2 / M_B^2$. Dashed and solid
lines are as in Fig. 1. The non-resonant contribution (a) and the sum of
non-resonant and resonant contributions (b) are separately shown.

\end{figure}

\newpage

\begin{figure}[htb]

\epsfxsize=16.5cm \epsfig{file=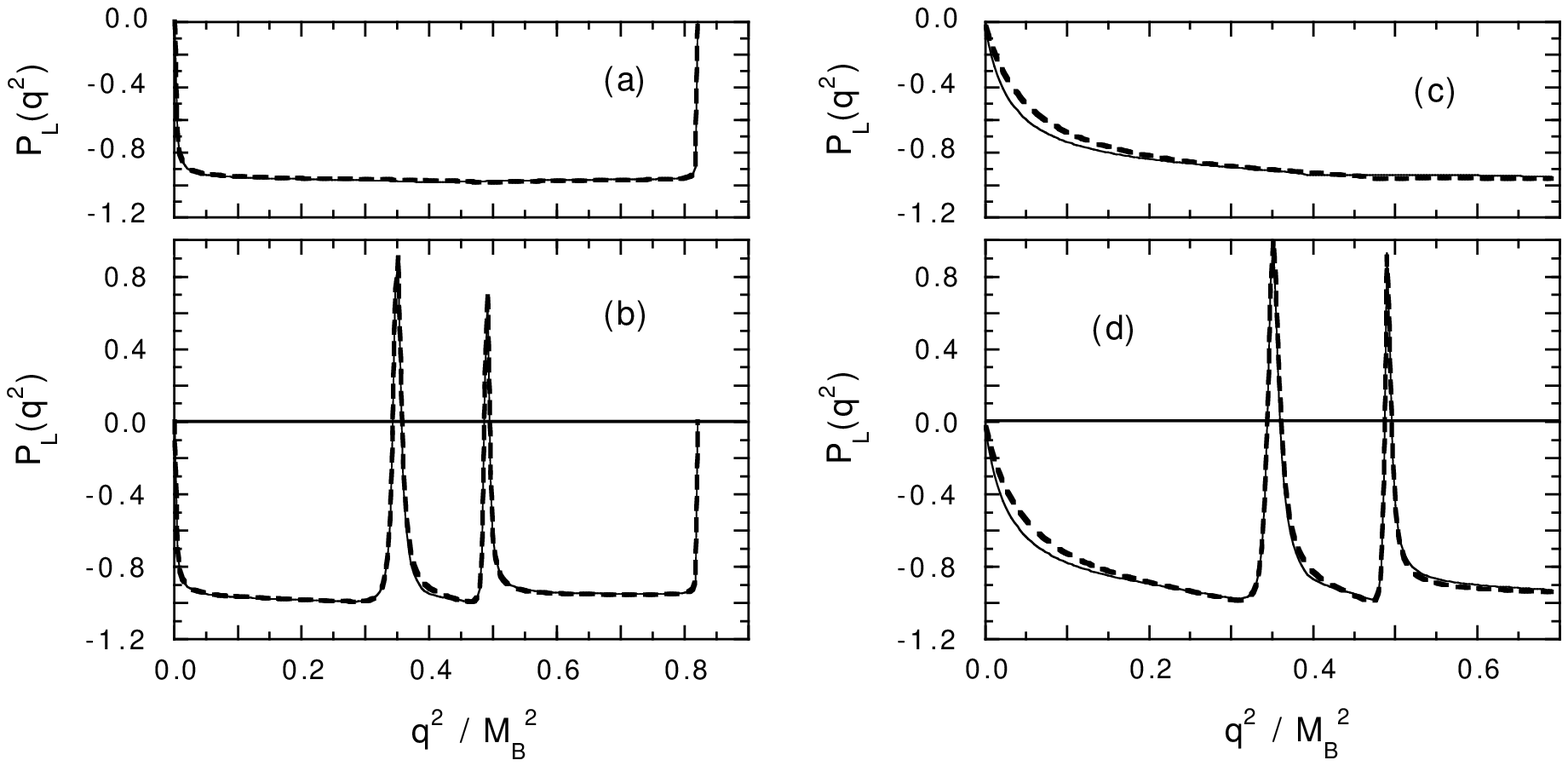}

\vspace{-15.5cm}

\noindent {\bf Figure 3.} The same as in Fig. 2, but for the lepton 
longitudinal polarization asymmetry ($P_L$) for the decays $B \to K \ell^+ 
\ell^-$ (a, b) and $B \to K^* \mu^+ \mu^-$ (c, d), respectively.

\end{figure}

}


\begin{thebibliography}{99}

\bibitem{gws} B. Grinstein, M.B. Wise and M.J. Savage: Nucl. Phys. {\bf
 B319} (1989) 271.

\bibitem{burasmuenz} A. Buras and M. M\"unz: Phys. Rev. {\bf D52} (1995)
 186.

\bibitem{ali} A. Ali, T. Mannel and T. Morozumi: Phys. Lett. {\bf B273}
 (1991) 505. A. Ali: Acta Phys. Pol. {\bf B27} (1996) 3529; Nucl. Instrum.
 Meth. {\bf A384} (1996) 8.  

\bibitem{rescontr} C.S. Lim, T. Morozumi and A.T. Sanda: Phys. Lett. {\bf
 B218} (1989) 343. P.J. O'Donnell and H.K.K. Tung: Phys. Rev. {\bf D43}
 (1991) R2067.

\bibitem{nunu} T. Inami and C. S. Lim: Prog. Theor. Phys. {\bf 65} (1981)
 287. G. Buchalla and A.J. Buras: Nucl. Phys. {\bf B400} (1993) 225.

\bibitem{iw} N. Isgur and M.B. Wise: Phys. Lett. {\bf B232} (1989) 113;
 Phys. Lett. {\bf B237} (1990) 527; Phys. Rev. {\bf D42} (1990) 2388.

\bibitem{jauswyler} W. Jaus and D. Wyler: Phys. Rev. {\bf D41} (1990)
 3405.

\bibitem{stech} B. Stech: Phys. Lett. {\bf B354} (1995) 447; Z. Phys.
 {\bf C75} (1997) 245.

\bibitem{burdman} G. Burdman: Phys. Rev. {\bf D52} (1995) 6400. W.
 Roberts: Phys. Rev. {\bf D54} (1996) 863.
 
\bibitem{colangelo} P. Colangelo {\em et al.}: Phys. Rev. {\bf D53} (1996)
 3672; Phys. Lett. {\bf B395} (1997) 339.

\bibitem{aliev} T.M. Aliev {\em et al.}: Phys. Lett. {\bf B400} (1997)
 194 and hep-ph 9612480.

\bibitem{gengkao} C.Q. Geng and C.P. Kao: Phys. Rev. {\bf D54} (1996)
 5636.

\bibitem{giw} C. Greub, A. Ioannissian and D. Wyler: Phys. Lett. {\bf
 B346} (1995) 349.

\bibitem{m} D. Melikhov: Phys. Rev. {\bf D53} (1996) 2460; Phys. Lett.
 {\bf B380} (1996) 363; Phys. Lett. {\bf B394} (1997) 385.

\bibitem{jaus} W. Jaus: Phys. Rev. {\bf D41} (1990) 3394; Phys. Rev. {\bf
 D53} (1996) 1349.

\bibitem{luke} M. Luke: Phys. Lett. {\bf B252} (1990) 447. M. Neubert and
 V. Rieckert: Nucl. Phys. {\bf B382} (1992) 97.

\bibitem{mn} D. Melikhov and N. Nikitin: hep-ph/9609503 (unpublished). 

\bibitem{isgw2} D. Scora and N. Isgur: Phys. Rev. {\bf D52} (1995) 2783.

\bibitem{gi} S. Godfrey and N. Isgur: Phys. Rev. {\bf D32} (1985) 189. 

\bibitem{sim95} F. Cardarelli {\em et al.}: Phys. Lett. {\bf B332} (1994)
 1; Phys. Lett. {\bf B349} (1995) 393; Phys. Lett. {\bf B359} (1995) 1;
 Few-Body Syst. Suppl. {\bf 9} (1995) 267; Phys. Rev. {\bf D53} (1996)
 6682.

\bibitem{sim96} S. Simula: Phys. Lett. {\bf B373} (1996) 193. I. Grach
 {\em et al.}: Phys. Lett. {\bf B385} (1996) 317.

\bibitem{cleo1} R. Ammar {\em et al.}: Phys. Rev. Lett. {\bf 71} (1993)
 674 and CLEO CONF 96-05 (1996).

\bibitem{cleo2} T. Skwarnicki: Proc. of the $17^{th}$ Int. Symp. on
 Lepton-Photon Interactions, Beijing (China), August 1995, eds. Z.
 Zhi-Peng and C. He-Sheng (World Scientific), p. 238.

\bibitem{cdf} C. Anway-Wiese {\em et al.}: FERMILAB CONF-95-201-E (1995).
 T. Speer {\em et al.}: FERMILAB CONF-96-320-E (1996). 

\end{thebibliography}
\end{document}